\documentclass[12pt]{iopart}

\usepackage{graphicx,bm}
\usepackage{amsfonts}
\begin{document}

\title{Quantum oscillation studies of the Fermi surface of iron-pnictide superconductors}

\author{A. Carrington}

\address{H.H. Wills Physics Laboratory, University of Bristol, Tyndall Avenue, Bristol, BS8 1TL UK.}
\ead{a.carrington@bristol.ac.uk}
\begin{abstract}
This paper reviews quantum oscillation studies of iron-pnictide superconductors and related materials.  These
measurements give unique information regarding the full three dimensional topology of the Fermi surfaces and the
renormalisation of the quasi-particle masses.  The review will cover measurements of the 122 arsenide end members,
XFe$_2$As$_2$ (X = Ba, Sr, Ca) which have a spin density wave ground state, but will concentrate on the phosphide end
members (LaFePO and XFe$_2$P$_2$) which have a paramagnetic ground state and a Fermi surface topology which is similar
to the higher $T_c$ superconducting iron-pnictides.  All three of the 122 phosphides become superconducting when P is
partially substituted by As, and for the BaFe$_2$(As$_{1-x}$P$_x$)$_2$ series de Haas-van Alphen oscillations are
observable for $0.42\leq x \leq 1$ with $T_c$ up to 25\,K.  The results show the changes in the Fermi surface topology
and the increase in the mass renormalisation as the correlations which induce superconductivity develop.
\end{abstract}


\section{Introduction}

A popular view of the iron-pnictide superconductors is that their unusual properties stem from the unique structure of
their Fermi surfaces. Very soon after the discovery of superconductivity at temperatures greater than 40\,K,
conventional density functional theory (DFT) band-structure calculations using the local density approximation (LDA)
(and its derivatives such as the generalised gradient approximation (GGA)) showed that the Fermi surface is mostly
composed of small, almost two dimensional, tubes running along the $c$-axis \cite{SinghD08}. The tubes at the centre of
the Brillouin zone are hole-like whereas those at the corner of the zone are electron-like [see Figs.\
(\ref{Fig:BandFold}), (\ref{Fig:LaFePO_Spag}) and (\ref{Fig:122_FS})]. For the undoped parent materials, for example
LaFePO or LaFeAsO, the volume of the hole and electron Fermi surfaces are exactly the same and the material is said to
be compensated as the number of electrons and holes are equal.

It was noticed that as these Fermi surfaces are both roughly circular in cross section and as they are the same size,
then a single translational vector $\bm{Q}=(\pi/a,\pi/a)$ maps one surface onto the other.  This property, known as
nesting, causes a peak in the Lindhard dielectric response function $\chi(\bm{q})$ at $\bm{q}=\bm{Q}$ which can lead to
charge or spin density wave instabilities depending on the relative strength of the electron-phonon or
electron-electron interactions. If the nesting is not so perfect, and assuming electron-electron interactions dominate
(as they do for iron-pnictides) then strong spin fluctuations should be expected with wavevectors centred around
$\bm{Q}$. In many models \cite{MazinSJD08,KurokiOAUTKA08,ChubukovEE08,CvetkovicT09} it is these spin-fluctuations which
provide the pairing interaction which binds the electrons into Cooper pairs and causes superconductivity. As the Fermi
surface geometry is at the heart of this interaction its experimental determination is of high importance.

It should be said before continuing that a completely itinerant electron model, as outlined above, is unlikely to be
the whole picture.  Although the many body electron-electron correlations are usually assumed to be relatively weak in
the iron-pnictides, as compared to the cuprate high $T_c$ superconductors for example, they cannot be neglected.  For
example, according to the DFT calculations the Fermi surfaces of LaFePO and LaFeAsO are very similar, however
experimentally LaFePO is a paramagnetic metal which becomes superconducting below 6\,K \cite{KamiharaHHKYKH06} whereas
LaFeAsO has a transition to spin density wave metallic antiferromagnetic state below 130\,K and is not superconducting
\cite{KamiharaWHH08}. An explanation for this difference is to suppose that the magnetic interactions originate
primarily from many body electron-electron correlations (for example the $U$ and $J$ parameters in Hubbard type models)
but the structure of the spin-fluctuations in momentum space is governed by the Fermi surface geometry \cite{MazinS09}.
This is backed up by the fact that DFT models generally correctly predict the form of the magnetic order in the static
SDW phase.  The relative importance of the local interactions and the itinerant behaviour is still under debate.

Given the importance of determining the properties of the Fermi surface it is not surprising that this has been the
subject of many theoretical and experimental investigations.  By far the most popular and ubiquitous experimental
technique is angle resolved photoemission spectroscopy (ARPES).  Here the sample is illuminated with photons, typically
of energy $6-100$\,eV, and emitted photoelectrons are collected and analysed.  As the in-plane momentum of the
electrons is conserved when they are emitted, measurements of the energy and momentum of these electrons as a function
of angle can be used to deduce the $\varepsilon(\bm{k})$ relationships for the various bands in the metal.  The
technique yields a direct `image' of a 2 dimensional slice through the Fermi surface at a particular $k_z$ which is
determined by the energy of the photons (with an uncertainty determined by the escape depth of the photoelectrons). One
problem with the technique is that it is surface sensitive. The surface electronic structure may differ from the bulk
for several reasons.  For example, the surface may be charged (and hence effectively doped) if the cleavage plane is
not neutral. The technique also has significantly lower resolution than quantum oscillations studies (particularly in
$k_z$).

\subsection{Determining the Fermi surface via quantum oscillations}

Measurements of the Fermi surface properties based on magneto-quantum oscillation (QO) effects are an important
complementary technique to ARPES.  Although the majority of the Fermi surface parameters which may be determined by the
QO effects are also accessible by ARPES, the main advantages of QO are: i) it is very insensitive to surface states as
it is dominated by the bulk response thus giving an unambiguous probe of the bulk Fermi surface, ii) it has very high
$\bm{k}$-space resolution in all three crystallographic directions - typically Fermi surface cross-sections can be
determined to better than $10^{-3}$ of the area of the Brillouin zone, iii) it has very high energy resolution, so that
it gives the effective mass very close to the Fermi level ($\Delta E\sim \mu_B B = $0.6\,meV for $B$=10\,T). A major
limiting factor is that only samples with sufficiently long mean free path will give a measurable signal. High magnetic
field and low temperatures are usually essential.

The technique is based on the fact that in a magnetic field the only allowed values of electron momentum $\bm{k}$ lie
on Landau tubes of area $A=(n+\frac{1}{2}) B/\phi_0$, where $\phi_0$ is the single electron flux quantum ($h/e$) and
$n$ is an integer known as the Landau index. These two dimensional electron tubes run parallel to the field $B$ and for
almost free electrons will have a circular cross-section.  As the field is varied the $\bm{k}$ space radius of a
particular Landau level $n$ will increase, and as each level passes through the Fermi surface the states will
depopulate giving rise to a saw-tooth like variation of the free energy, periodic in inverse field $1/B$. The frequency
(in inverse field) of the oscillations is related to the $\bm{k}$-space cross sectional area of the Fermi surface
$A_{\bm k}$ by the Onsager relation, $F=\hbar A_{\bm k}/2\pi e$.  For a three dimensional metal $A_{\bm k}$ will vary
as a function of $k_z$ and hence each $k_z$ slice (assuming $B\| z$) will give rise to oscillations of varying
frequency and phase. The total signal will be the sum of the contributions from each slice. Usually the sum is
dominated by the extremal cross-sections of the Fermi surface where $dA_{\bm k}/dk_z=0$ and hence a typical Fermi
surface sheet will give rise to a small number of oscillatory signals.

The oscillations in the free energy may be determined by measuring many different physical properties such as the
resistivity (Shubnikov-de Haas effect), the magnetisation or magnetic torque (de Haas-van Alphen effect), specific heat
etc.  Finite temperature, impurity scattering and other effects smear out the discontinuity as the Landau levels pass
through $\varepsilon_F$. The response may be decomposed into a Fourier series, and the signal is usually dominated by
the fundamental harmonic at frequency F.  For a three dimensional metal of volume $V$ the full expression for the first
harmonic of the oscillatory torque $\mathcal{T}$ is (in SI units)
\begin{eqnarray}
\mathcal{T} &=&\sum_{\rm orbits}\frac{\partial F}{\partial \theta}  \frac{Ve^{5/2}}{\hbar^{1/2}\pi^2 m_e} \frac{B^{3/2}}{\left(2\pi \frac{\partial^2 A}{\partial k^2_\|}\right)^{1/2}}R_T R_D R_S \sin\left(\frac{2\pi F}{B} + \phi\right),\\
R_D &=& \exp\left(-\frac{\pi m_b}{eB\tau}\right), \nonumber\\
R_T &=& \frac{X}{\sinh X}, \quad X=\frac{2\pi^2k_BTm^*}{e\hbar B}, \nonumber\\
R_S &=& \cos \left(\frac{\pi gm_S}{2m_e}\right).\nonumber
\label{Eq:LK}
\end{eqnarray}

This is commonly known as the Lifshitz-Kosevich (LK) formula \cite{Shoenberg}.  The expression includes the most common
attenuation factors, $R_T$, $R_D$ and $R_S$ which are due to finite temperature, impurity scattering and spin-splitting
respectively.  It should be noted that there are other attenuation factors, due to effects such as mosaic crystal
structure or superconductivity which are important in some circumstances.  In the expression the quasiparticle mass
occurs several times in different forms.  $m_e$ is the free electron mass and $m_b$ is the band mass as calculated from
mean-field DFT band structure calculations.  The mass $m^*$ which occurs in the expression for the temperature
dependent damping factor $R_T$ is the quasiparticle effective mass renormalized by electron-phonon and
electron-electron interactions. $m_s$ is the 'spin-mass' which is renormalized by electron-electron interactions (which
enhance the Pauli susceptibility) only. The curvature factor, $(\frac{\partial^2 A}{\partial k^2_\|})^{-1/2}$ accounts
for the number of states which contribute to the extremal orbit, here $k_\|$ is the component of $\bm{k}$ parallel to
the field. If $A_{\bm k}$ changes little with $k_\|$, for example because the Fermi surface is locally almost two
dimensional, then these orbits will have a large intrinsic amplitude.

The Dingle factor, $R_D$, is determined by the impurity scattering rate $\tau^{-1}$ and is not normally renormalized by
many body factors. For free electrons we can make the substitutions, $m_b v_F = \hbar k_F$, and $\ell=\tau v_F$, where
$v_F$ is the Fermi velocity and $\ell$ is the mean free path, so that
\begin{equation}
R_D = \exp\left(-\frac{\pi \hbar k_F}{eB\ell}\right)
\label{Eq:Dingle}
\end{equation}
For the general case, $\ell$ and $k_F$ should be replaced by values averaged over the orbit.  A simple approximation is
to take $\pi k_F^2=A=2\pi e F /\hbar$, so that $R_D=\exp(-1140 \sqrt{F}/\ell B)$, with $\ell$ in Angstroms and $F,B$ in
Tesla.

This paper will review QO experimental studies of the Fermi surface of the iron-pnictide superconductors. Because of
the need for high purity samples these studies have mostly been conducted on the stoichiometric end members of the
various iron-pnictide series.  Charge doping of these parent materials, which is a common route to induce
superconductivity, generally induces significant disorder which shortens the mean-free-path to such an extent that QO
are not observable even in the highest available fields ($B_{\rm max}\simeq 70$\,T).  On the other hand, isovalent
substitution of As by P in the series BaFe$_2$(As$_{1-x}$P$_x$)$_2$ does not induce so much disorder and QO are
observed for dopings producing $T_c$ up to $\sim$80\% of the maximum \cite{Shishido10}.

\section{Parent iron-arsenides in SDW state}
In their undoped state, at ambient pressure,  the `1111' (e.g., LaFeAsO) and `122' (e.g., BaFe$_2$As$_2$)
iron-arsenides have a spin-density wave ground state.  Below a temperature of 140\,K BaFe$_2$As$_2$ undergoes a
structural and magnetic transition (at slightly different temperatures) to a state with stripe ordered magnetic moments
with a unit cell which is roughly $\sqrt{2}$ larger (along the edge) and rotated by 45$^\circ$ compared to the high
temperature cell \cite{Rotter08}.  The structure is actually orthorhombic (\textit{Fmmm}), with \textit{a}=5.6146\,\AA,
\textit{b}=5.5742\,\AA,and \textit{c}=12.9453\,\AA. The fact that \textit{a} and \textit{b} only differ by 0.7\% means
that the crystals are highly susceptible to twinning, so that crystals usually have multiple domains in which the
\textit{a} and \textit{b} directions are switched. Twin free samples have been produced by cooling through the SDW
transition with an applied uniaxial stress or a large magnetic field \cite{Tanatar10,ChuAPDLYF10,ChuADMIYF10}.

\begin{figure*}
\flushright
\includegraphics*[width=13cm]{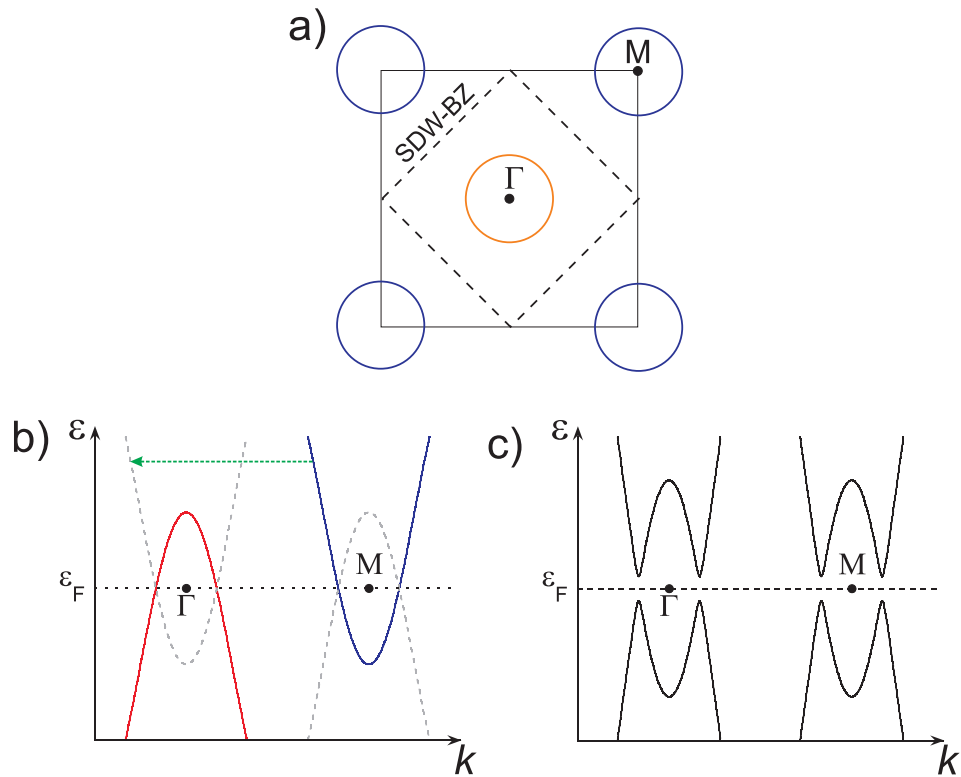}
\caption{Simplified schematic picture of Fermi surface reconstruction due to band folding at the spin-density wave
transition in the iron-pnictides.  a) Shows circular Fermi surface sheets, hole like at the zone centre ($\Gamma$) and
electron like at the zone corner ($M$), in the paramagnetic Brillouin zone (solid line). The change in structure at the
SDW transition will cause the Brillouin zone to change (dashed line). b) shows the band-structure energy - momentum
curves corresponding to the Fermi surface in a).  As the Brillouin zone changes the band at $M$ is translated (folded)
back to the $\Gamma$ point in the reduced zone scheme.  c) The folded bands hybridize and a gap forms at the Fermi
level $\varepsilon_F$, and the Fermi surface disappears.} \label{Fig:BandFold}
\end{figure*}

As a first approximation the effect of this change in structure on the Fermi surface can be understood by a simple
band-folding argument. As illustrated in Fig.\ \ref{Fig:BandFold}, the change in structure causes the electron band
which produced a Fermi surface at the corner of the original cell to fold to the centre of the zone.  The electron-like
bands and the hole-like bands now cross and in the simple case where the shape and size of hole and electron Fermi
surfaces are identical, the bands would cross exactly at the Fermi level. The bands will then hybridize and a gap will
appear at the Fermi level, hence the Fermi surface would completely disappear and the material would become insulating.
In reality, the original Fermi surface sheets do not have exactly the same in-plane shape and there is some corrugation
along $k_z$, hence the Fermi surface will not be completely wiped out and small three dimensional Fermi pockets will
remain.  In addition, the formation of the SDW state causes the bands to shift in energy (some bands more than others)
so the exact band-structure is not simply related to the folded paramagnetic structure.  Fixed moment DFT calculations
seem to produce approximately the correct structure - although not the exact Fermi surface topology \cite{Yi2009}.

\begin{figure*}
\flushright
\includegraphics*[width=13cm]{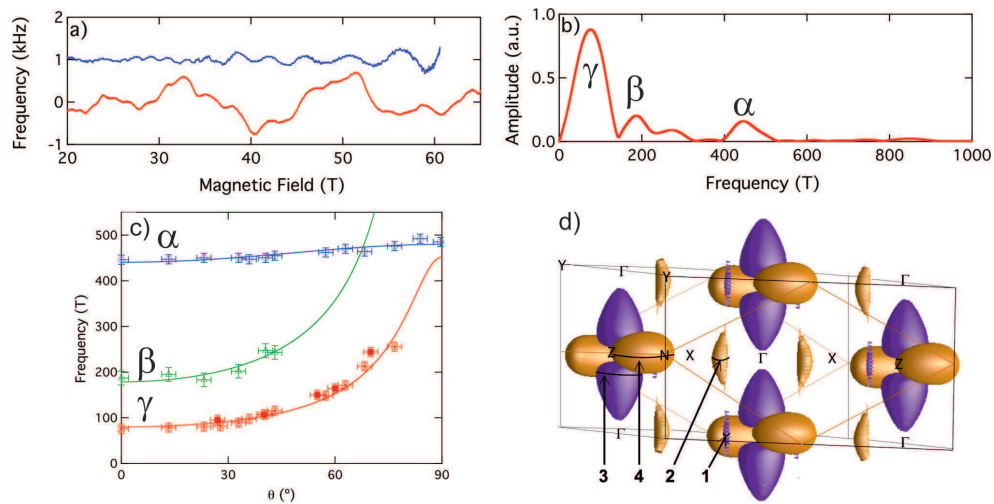}
\caption{Quantum oscillations in BaFe$_2$As$_2$. a) Raw oscillator frequency shift which is proportional to the
resistance versus field for $B\|c$ and $B\perp c$ (upper and lower curves respectively), b) Fourier transform of the
$B\|c$ data. c) Variation of observed SdH peaks with field angle (solid lines are fits to a simple elliptical Fermi
surface model). d) DFT fixed moment calculation of the Fermi surface using the negative $U$ procedure to fix the
ordered moment.  The predicted extremal orbits are labelled 1-4. Figure adapted from Analytis \emph{et
al.}\cite{AnalytisMCRBKJF09}.} \label{Fig:BaFe2As2}
\end{figure*}

Quantum oscillation experiments have been reported on several members of the `122' family XFe$_2$As$_2$ with X=Sr, Ba
and Ca \cite{SebastianGHLSML08,AnalytisMCRBKJF09,HarrisonMMBRT09}. Examples of the raw (background subtracted)
experimental results for the oscillations in the resistivity of BaFe$_2$As$_2$(measured inductively using a tunnel
diode oscillator) are shown in Fig.\ \ref{Fig:BaFe2As2}.  The signals are relatively weak and can only be observed in
very high magnetic field (typically greater than 30\,T). In all these measurements only very low oscillation
frequencies were observed.  For example for BaFe$_2$As$_2$, the values with $B\|c$ were reported to be 80\,T, 190\,T
and 440\,T, which correspond to 0.3, 0.7 and 1.7\% of the total area of the paramagnetic Brillouin zone respectively
\cite{harmonic}.The weak variation of the $\alpha$ frequency with magnetic field angle shows that this comes from a
small quasi spherical pocket as anticipated. A DFT calculation of the Fermi surface of BaFe$_2$As$_2$ based on the
observed magnetically ordered unit cell is shown in Fig.\ \ref{Fig:BaFe2As2}c). These calculations use the LDA+U
procedure, which is common for strongly correlated metals, with a negative orbital potential $U$ to reduce the ordered
moment. This negative $U$ has no physical significance and should be regarded as an empirical fix to the LDA
calculations. With moderate shifts of the calculated bands energies ($<60$\,meV), the predicted orbits 1-3 can be
brought into agreement with the experimental results in Ref.\ \cite{AnalytisMCRBKJF09}. In this study the Fermi surface
determination was incomplete as all the predicted orbits were not observed experimentally.

Recently a complete Fermi surface determination of de-twinned BaFe$_2$As$_2$ was reported \cite{Terashima1103.3329}. In
these detwinned samples the oscillations are large in amplitude even in low fields less than 10\,T.  All extremal
orbits of the calculated Fermi surface were observed, and the calculations could be brought into almost exact agreement
with experiment with appropriate band energy shifts ($<65$\,meV).  The determined Fermi-surface is quite similar that
shown in Fig.\ \ref{Fig:BaFe2As2} but has some differences in detail for some of the pockets.

\section{LaFePO}
The 1111 compound LaFePO was the first discovered iron-pnictide superconductor \cite{KamiharaHHKYKH06} and was also the
first paramagnetic iron-pnictide in which QO were observed \cite{ColdeaFCABCEFHM08}.  Unlike the isomorphic arsenide
LaFeAsO, LaFePO is superconducting in its stoichiometric state (some reports have suggested that a few percent of
oxygen vacancies are needed to induce superconductivity \cite{Hamlin08,McqueenRWHLHWGC08}, however, it is not clear if
this is necessary \cite{ColdeaFCABCEFHM08}). Unlike LaFeAsO,  LaFePO is not magnetically ordered and magnetism can not
be induced by small amounts of doping away from stoichiometry.  The superconducting transition temperature is quite low
$T_c \simeq 6$\,K and the upper critical fields ($H_{c2}\|c=0.68$\,T,  $H_{c2}\perp c=7.2$\,T) are easily exceeded
with conventional superconducting magnets \cite{CarringtonCFHABACEFM09}. It is therefore an ideal candidate for study
by quantum oscillations effects.

\begin{figure*}
\flushright
\includegraphics*[width=12cm]{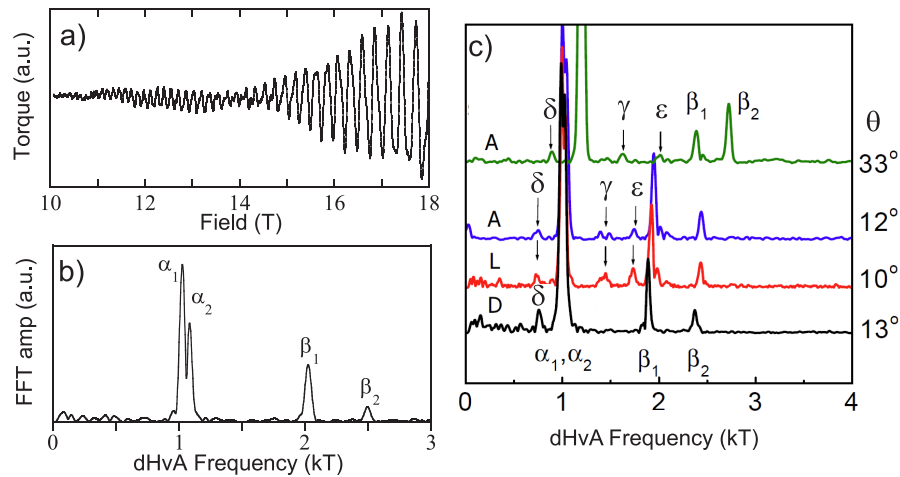}
\caption{Quantum oscillations in LaFePO. a) Raw torque versus field at $T$=0.35\,K with $4^{th}$ order polynomial
subtracted. b) Fast Fourier transform of the same data. c) FFT spectra of different samples (labelled A, L D) at
several angles $\theta$ showing the various extremal orbit peaks.  Figure adapted from data in Ref.\
\cite{ColdeaFCABCEFHM08}.} \label{Fig:LaFePO_Raw}
\end{figure*}

\begin{figure*}
\flushright
\includegraphics*[width=13cm]{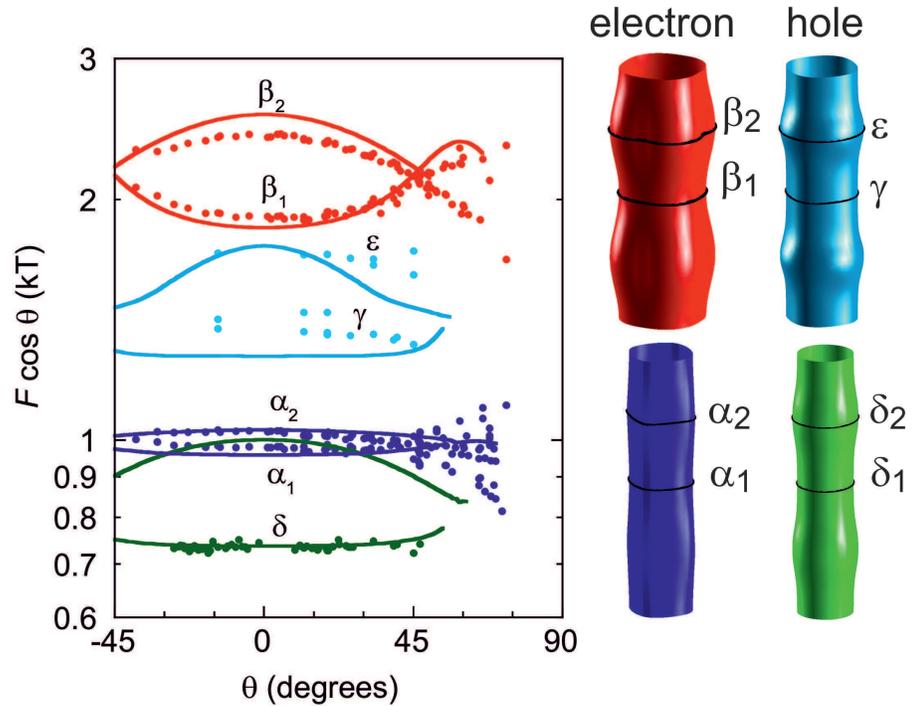}
\caption{Observed dHvA oscillation frequencies in LaFePO (solid symbols) versus field angle as the field is rotated
from $B\|c$ ($\theta=0^\circ$) to $B \perp c$.  The values derived from the DFT calculation, after shifting the band
energies (as described in the text) are shown by the solid lines. The right panel shows the extremal orbits on the
electron and hole Fermi surface sheets. Figure adapted from data in Carrington \textit{et al}
\cite{CarringtonCFHABACEFM09}.} \label{Fig:LaFePO_Rotation}
\end{figure*}

Data were first reported by Coldea \textit{et al} \cite{ColdeaFCABCEFHM08}, closely followed by Sugawara \textit{et
al.}\cite{SugawaraSDMKYO08}  As the best available single crystals of LaFePO are quite small (typically planar
dimensions of a few hundred microns and thickness a few tens of microns) the microcantilever torque method is well
suited.  Examples of raw torque versus magnetic field data for this compound are shown in Fig.\ \ref{Fig:LaFePO_Raw}.
The data are best appreciated by calculating the fast Fourier transform (with respect to inverse field). A total of 7
fundamental frequencies are observed plus harmonics of the strongest amplitude signals ($\alpha$ and $\beta$ orbits).
Measurements were repeated as a function of field angle as the field was rotated from parallel to the $c$-axis
($\theta=0^\circ$) towards the $ab$-plane.   As the field is rotated the extremal electron orbits transverse different
sections of the Fermi surface, and hence by analyzing the field angle dependence of the dHvA frequencies, $F(\theta)$,
a detailed model of the Fermi surface can be constructed.

For a perfectly two dimensional Fermi surface, $F$ is inversely proportional to $\cos \theta$ and hence by plotting
$F\cos \theta$ versus $\theta$ small deviations from two dimensionality can be easily seen. The simplest distortion of
a two dimensional Fermi surface in a tetragonal system is a cosine dispersion, where the $\varepsilon(k) = \hbar^2
(k_x^2+k_y^2)/2m+t_\perp \cos(ck_z)$.  In this case, there are two extremal frequencies, the minimum $F_{\rm min}$ and
the maximum $F_{\rm max}$, which vary with angle according to the Yamaji formula \cite{Yamaji89}

\[
F_{\rm min,max}(\theta) \cos\theta = F_{av} \pm \Delta F J_0(c k_F \tan\theta).
\]
In this expression, the first term is the mean frequency ($F_{av}=F_{\rm min}+F_{\rm max})/2$ at $\theta=0$, the
prefactor of the second term is $\Delta F = (F_{\rm max}^0-F_{\rm min}^0)/2$,  $J_{0}$ is the Bessel function and
$k^2_F=2eF_{av}/\hbar$. At certain angles $J_0=0$ and $F_{\rm min}=F_{\rm max}$ and here all parts of the Fermi surface
interfere constructively to give a large increase in the amplitude of the oscillations. Such a point is observed at
$\theta =42(2)^\circ$ for the $\beta$ frequencies of LaFePO, which together with the overall behaviour of $F_{\rm min}$
and $F_{\rm max}$ are very strong evidence that they originate from a single $c$-axis warped section of Fermi surface
(see Fig.\ \ref{Fig:LaFePO_Rotation}). The two $\alpha$ frequencies are much closer indicating that they come from an
almost two dimensional section of Fermi surface with little warping, and no clear Yamaji amplitude peak is seen.  For
the lowest frequency $\delta$, $F\cos\theta$ shows almost no change with $\theta$, consistent with it coming from an
almost 2D Fermi surface section. The remaining two frequencies $\varepsilon$ and $\gamma$ are much weaker; they were
only observed above $B$=30\,T and again $F\cos\theta$  shows minimal change with $\theta$ within the noise.

\begin{figure*}
\flushright
\includegraphics*[width=13cm]{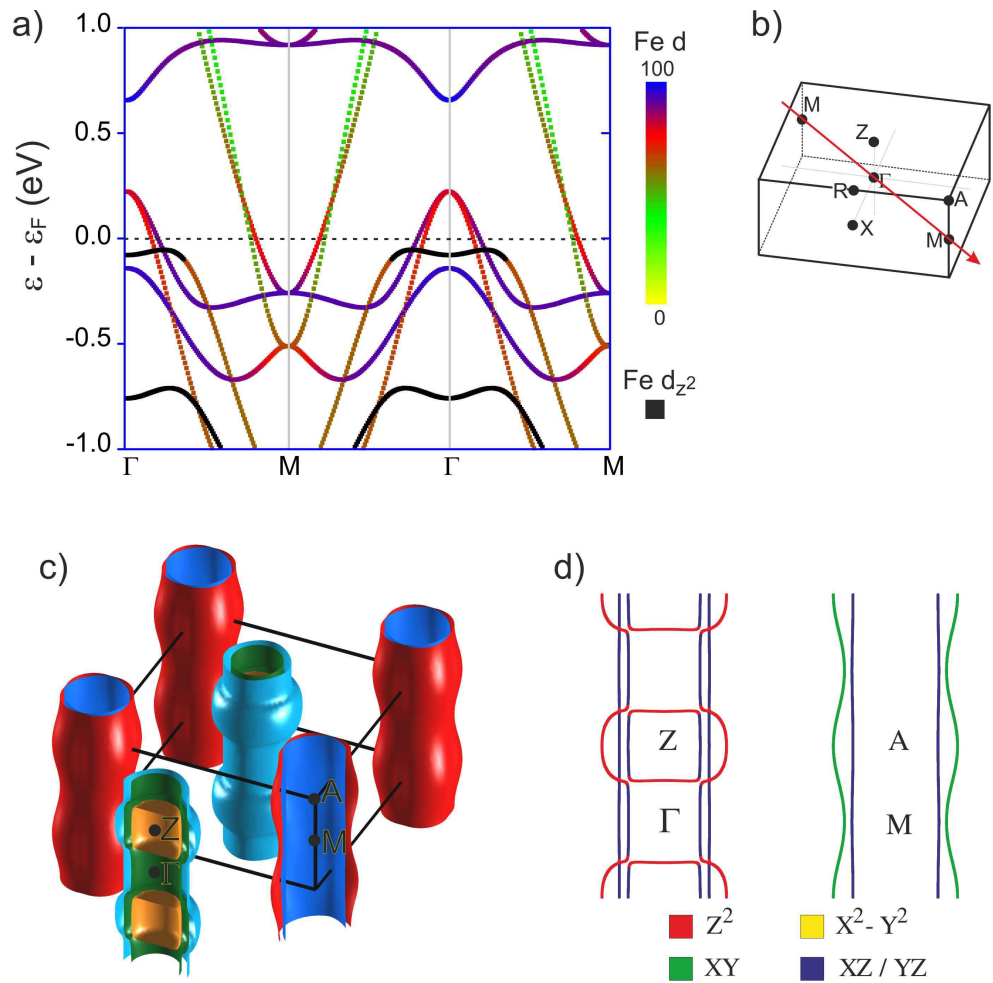}
\caption{Calculated electronic structure of LaFePO without spin orbit interaction or adjustment to fit experimental
data.  a) Band structure energy-momentum along the zone diagonal. The percentage of Fe-$d$ character is indicated by
the colour scale, with the band with predominant Fe $d_{z^2}$ character overprinted in black. b) Brillouin zone with
symmetry labels. c) Three dimensional Fermi surface. d) $c$ axis (110) slice through the Fermi surface with predominant
band character indicated by the colour scale.  Figure adapted from Ref.\ \cite{CarringtonCFHABACEFM09}.}
\label{Fig:LaFePO_Spag}
\end{figure*}

\begin{figure*}
\flushright
\includegraphics*[width=13cm]{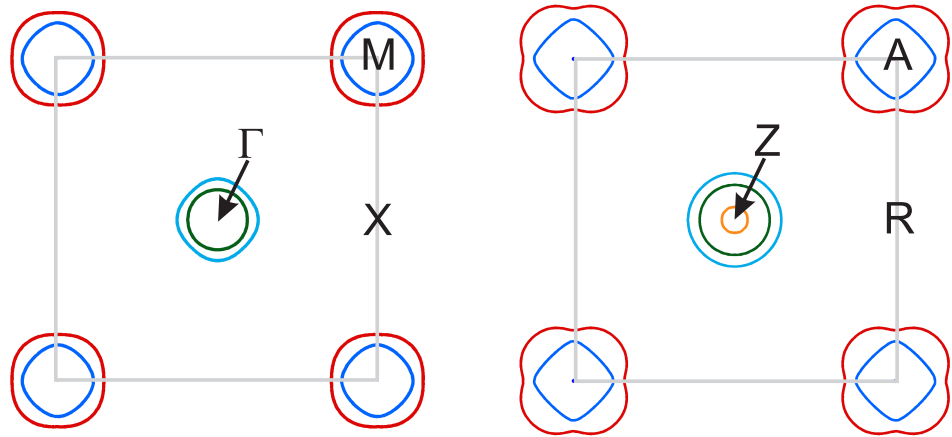}
\caption{In plane cross sections of the calculated Fermi surface of LaFePO including the spin-orbit interaction and
with bands shifted to best match the dHvA data. Figure adapted from Ref.\ \cite{CarringtonCFHABACEFM09}.}
\label{Fig:LaFePO_Slices}
\end{figure*}

Further analysis of the data is greatly aided by band-structure calculations. The experimental dHvA frequency versus
field angle data can be compared to these calculations and adjustments made to best fit the data.  Often this can be
accomplished with a small number of parameters, usually a single energy shift for each band or group of bands. The
alternative is to fit the data to a series of appropriate geometrical harmonics.  This latter approach works well when
the Fermi surface is relatively simple and only has small deviations from a sphere or a two dimensional tube and only a
small number of harmonics are required (for example, SrRu$_2$O$_4$ \cite{BergemannMJFO03} or
Tl$_2$Ba$_2$CuO$_{6+\delta}$  \cite{BanguraRBMCHC10,RourkeBBMCCH10}).

For LaFePO the band-structure calculations were done \cite{ColdeaFCABCEFHM08,CarringtonCFHABACEFM09} using the
augmented plane wave plus local orbitals methods as implemented in the WIEN2K code \cite{wien2k}. For these
calculations the experimental lattice parameters and internal coordinates were used.  A plot of the energy-momentum
band structure along the $\Gamma-M$ direction (Fig.\ \ref{Fig:LaFePO_Spag}) shows two hole bands crossing the Fermi
level close to $\Gamma$ and two electron bands crossing close to $M$. So apart from the doubling of bands the detailed
band-structure looks quite similar to the simplified picture in Fig.\ \ref{Fig:BandFold}.  There are however two
complications.  First the outer electron surface has significant $c$-axis and in-plane warping and second, there is a
fifth band, mostly coming from the Fe $d_{z^2}$ orbital which cuts across the two dimensional hole bands and forms a
closed three dimensional pocket close to $Z$.

The assignment of the  $\alpha$ and $\beta$ orbits to the inner and outer electron sheets can be made rather easily.
The calculation fairly accurately predicts the size and warping of these sheets. The relatively strong $c$-axis warping
of the $\beta$ electron sheet and the much more two dimensional nature of the $\alpha$ electron sheet is well
reproduced. To get exact agreement with the data it is necessary to shift the calculated electron bands up in energy
(+85\,meV and +30\,meV for the $\alpha$ and $\beta$ orbits respectively).  The physical meaning of these band shifts
will be discussed later. With this adjustment the agreement is very good. The calculated $c$-axis warping of the
$\beta$ sheet is slightly too large: experimentally $\Delta F/F_{av}$=0.23 whereas the calculation gives $\Delta
F/F_{av}$=0.33, but for the $\alpha$ sheet it is much closer.  Without the inclusion of spin-orbit coupling the two
electron bands are degenerate along the zone edges $X-M$ and $R-A$. This degeneracy is lifted by the spin-orbit
interaction and a gap of $\sim 50$\,meV appears in the calculations. As the two electron bands need to be shifted by
different amounts to get agreement with experiment this implies that this gap is actually about double this (i.e.,
around 100\,meV).   This gap can be seen on the cross sections through the fitted Fermi surfaces shown in Fig.\
\ref{Fig:LaFePO_Slices}.

Identification of the hole orbits is difficult.  The lowest observed frequency $\delta$ is likely the minimum of the
inner hole sheet which is centred at $\Gamma$.  As predicted by the calculation there is minimal change in
$F\cos\theta$ for this orbit as the sheet is locally close to two dimensional at this point.  It is brought into
perfect alignment by shifting all the hole bands down by 53\,meV.  In that case the maximum of the inner hole band
almost perfectly coincides with the inner electron band $\alpha$.  The assignment of the other two frequencies $\gamma$
and $\varepsilon$ is more problematic. With the above mentioned shift of the three hole band energies it can be seen
(Fig.\ \ref{Fig:LaFePO_Rotation}) that $\gamma$ and $\varepsilon$ approximately correspond to the minimum and maximum
of the largest hole sheet.  The variation of $\varepsilon$ with $\theta$ however is much less than expected. As these
orbits are only observed at high magnetic field ($>$30\,T) it is possible that either or both of them result from
 magnetic breakdown or torque interaction effects arising from the electron orbits \cite{CarringtonCFHABACEFM09}.
In the calculations, the energy of the Fe $d_{z^2}$ band, which gives rise to the warping of the hole sheets and the
small spherical pockets at $Z$, varies very strongly with the height of the P atom and so it is possible that it does
not actually cross the Fermi level. In this case the above rigid band shifting will not work accurately because the
$d_{z^2}$ band has crossed and hybridized with the two dimensional $d_{xz\|xy}$ bands.  Instead, we might expect the
hole bands are almost perfectly two dimensional, thus giving rise to only two dHvA frequencies with no variation in
$F\cos\theta$ as a function of $\theta$.  $\delta$ is almost certainly from the inner hole band but which of $\gamma$
and $\varepsilon$ is from the largest hole band is uncertain.

The effective mass of the quasiparticles on each of the orbits was deduced by measuring the temperature dependence of
the dHvA oscillation amplitudes and fitting this to the expression for $R_T$ (Eq.\ (1)).  The masses for the electron
orbits range from 1.7 to 1.9 $m_e$. This should be compared to the masses calculated from the band-structure (for the
shifted bands) which range from 0.7 to 0.9 $m_e$, hence the many-body mass enhancement factors $m^*/m_b-1=\lambda$
range from 1.0(2) to 1.7(2) for the 4 orbits (see Table \ref{Table:massenhacements}).  The correlation is much stronger
than expected from electron-phonon scattering $\lambda_{ep}\simeq 0.21$ \cite{Boeri08}, and likely results from
electron-electron interactions.  ARPES measurements on LaFePO concluded that the bands were all narrowed by a factor
$\sim 2.2$, hence this implies that the electron correlations act over a wide energy range rather than being localised
close to the Fermi level.  This goes against interactions with low energy spin fluctuations being the major cause of
the mass enhancement. The masses of the hole orbits are quite similar, however the errors are larger on account of the
smallness of the signals.  It is unclear at present whether the masses of the observed orbits are sufficient to account
for the measured electron specific heat $\gamma_e$.  Four two dimensional Fermi surfaces each with $m^*/m_e=2$ would
give $\gamma_e=6$\,mJ mol$^{-1}$ K$^{-2}$ whereas the reported value on polycrystalline samples was $\gamma_e=12.7$\,mJ
mol$^{-1}$ K$^{-2}$ \cite{Hamlin08}. This discrepancy could indicate that a heavy hole sheet ($m^*/m_e \simeq 11$) has
not yet been detected but the heat capacity measurement also needs to be checked on \textit{single} crystal samples.

\section{122 phosphides: BaFe$_2$P$_2$, SrFe$_2$P$_2$, CaFe$_2$P$_2$}

Studies of the Fermi surfaces of the 122 phosphides are important for a number of reasons. Unlike the corresponding
arsenides, the Fermi surfaces of the phosphides are very similar to those of the superconducting phases of the Ba and
Sr 122 materials. In all three (Ba, Sr, Ca) phosphides superconductivity can be induced by partial substitution of P
with isovalent As \cite{Jiang09,Shi10} and hence the pure phosphide can be viewed as the end member of these series.
The maximum $T_c$ produced by P/As substitution is very comparable to that obtained by the non-isovalent dopants K
\cite{Rotter08} / Co \cite{sefat08} or pressure \cite{AlirezaKGPCLS09}.

\begin{figure*}
\flushright
\includegraphics*[width=13cm]{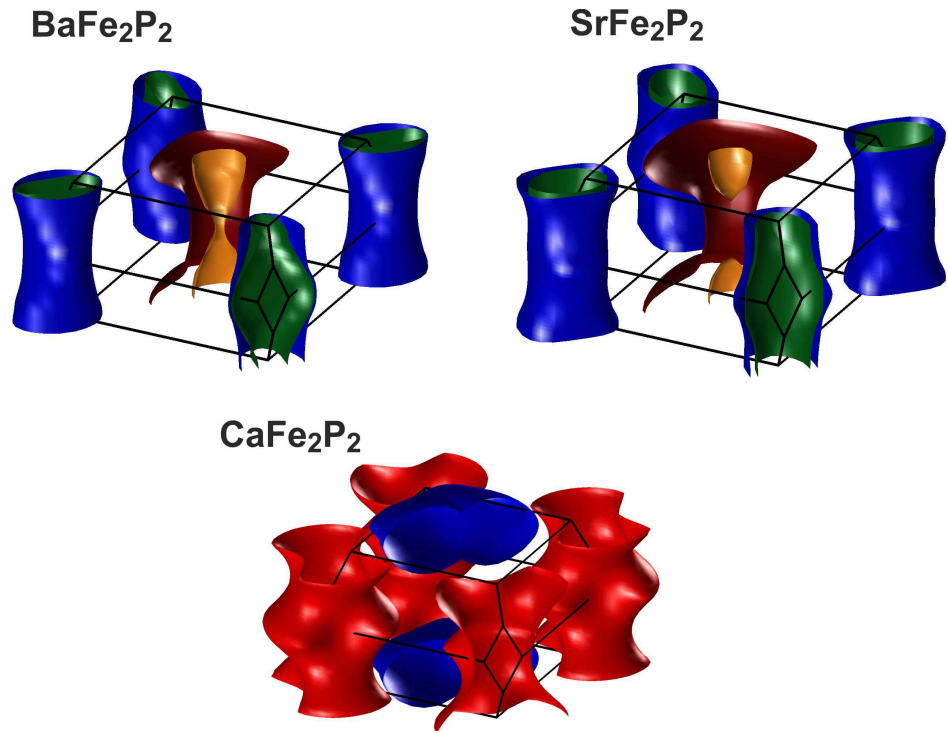}
\caption{Fermi surfaces of the 122 iron phosphides, BaFe$_2$P$_2$, SrFe$_2$P$_2$, and  CaFe$_2$P$_2$, determined by DFT
calculations with bands shifted to best match the dHvA measurements \cite{AnalytisACMCMFC09,Coldea09,Arnold11}.}
\label{Fig:122_FS}
\end{figure*}

The 122 phosphides are all paramagnetic, non-superconducting metals.  Calculations of the Fermi surface of the Ba and
Sr phosphides closely resemble those of the corresponding arsenides in their paramagnetic state (i.e., above $T_{\rm
SDW}$). The Fermi surface of CaFe$_2$P$_2$ on the other hand, closely resembles that calculated for CaFe$_2$As$_2$ in
its high pressure state.  One effect of replacing As by P in the 122 materials is to decrease the distance between the
Fe planes and hence increase the pnictogen-pnictogen bonding. In the case of CaFe$_2$As$_2$ the application of pressure
also decreases the interplane distance and for $P>0.35$\,GPa there is sudden, very large ($\sim$10\%) decrease in the
$c$-axis length. This high pressure phase is known as the collapsed tetragonal state.  CaFe$_2$P$_2$ has a similar
interplane distance as the collapsed tetragonal state of CaFe$_2$As$_2$ and hence a similar Fermi surface.  The
calculated Fermi surfaces for all three compounds are shown in Fig.\ \ref{Fig:122_FS}.

\begin{figure*}
\flushright
\includegraphics*[width=13cm]{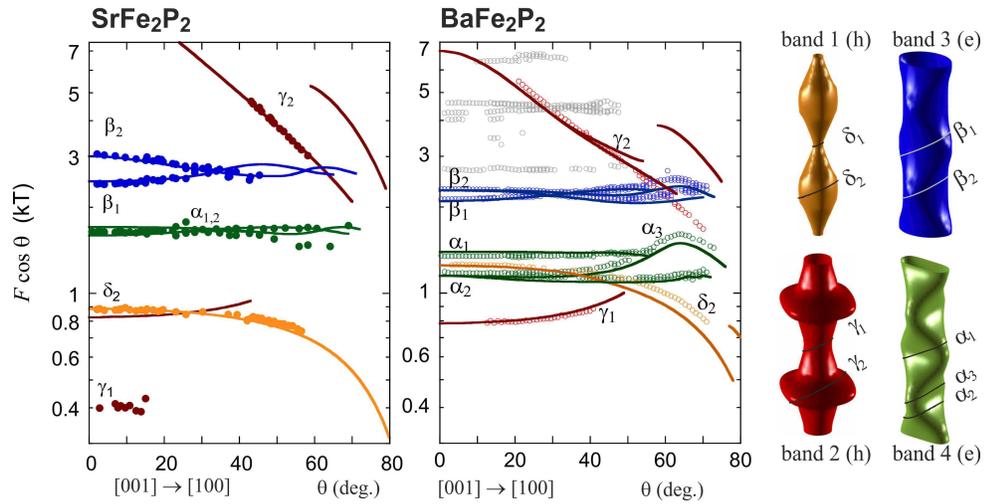}
\caption{Experimental dHvA frequencies versus magnetic field angle for rotations from $B\|c$ [001] to $B\|a$ [100] for
both SrFe$_2$P$_2$ \cite{AnalytisACMCMFC09} and BaFe$_2$P$_2$ \cite{Arnold11} (symbols).  The unlabelled points in grey
in the BaFe$_2$P$_2$ panel are harmonics of the $\alpha$ and $\beta$ orbits.  The orbit labels for SrFe$_2$P$_2$ have
been changed from that in Ref.\ \cite{AnalytisACMCMFC09} to be consistent with those used for BaFe$_2$P$_2$.  The solid
lines are the frequencies calculated from the DFT calculations with shifted bands. The far right panel depicts the
locations of the various orbits on the Fermi surface of BaFe$_2$P$_2$. The labels also apply to SrFe$_2$P$_2$ which has
a very similar Fermi surface (see Fig.\ \ref{Fig:122_FS}). Figure adapted from Refs.\ \cite{AnalytisACMCMFC09} and
\cite{Arnold11}} \label{Fig:BaSrFe2P2_Rot}
\end{figure*}
\begin{figure*}
\flushright
\includegraphics*[width=13cm]{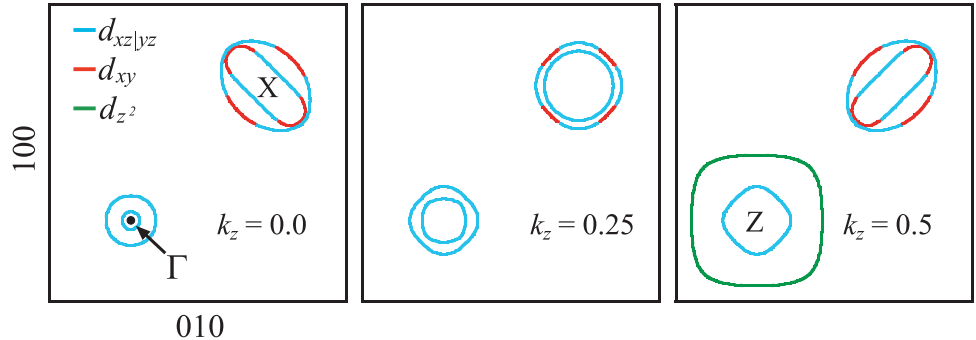}
\caption{Cross-sections of the determined Fermi surface of BaFe$_2$P$_2$ in the $ab$ plane at three different $k_z$
values (quoted in units of $c^*$). The strongest band character at each $k$-point is indicated. Note that the zone
corner is labelled X in the body centred unit cell of the 122 compounds.  This is equivalent to M in the simple
tetragonal unit cell of LaFePO.  Figure adapted from Ref.\ \cite{Arnold11}.} \label{Fig:BaFe2P2_Slices}
\end{figure*}


dHvA studies have been completed for all three of these 122 phosphides
\cite{AnalytisACMCMFC09,Coldea09,Shishido10,Arnold11}. The oscillations can be observed down to low field ($\sim 5$\,T)
and clear dHvA peaks in the Fourier spectra are observed for orbits on all four sheets of Fermi surface. Experimental
dHvA frequencies versus magnetic field angle for both SrFe$_2$P$_2$ \cite{AnalytisACMCMFC09} and BaFe$_2$P$_2$
\cite{Arnold11} are shown in Fig.\ \ref{Fig:BaSrFe2P2_Rot}.  For SrFe$_2$P$_2$ by far the largest signal comes from the
$\alpha$ (inner electron) orbits, followed by the $\beta$ (outer electron) orbits. The large $\alpha$ orbit signal
originates from the fact that the in-plane area of this orbit changes very little with $k_z$ and hence the curvature
factor (Eq.\ \ref{Eq:LK}) is large. This is demonstrated by the flat behaviour of $F_\alpha\cos(\theta)$ in Fig.\
\ref{Fig:BaSrFe2P2_Rot}. It should be noted however, that in this case, this does \textit{not} mean that this sheet of
Fermi surface is \textit{highly} two dimensional.  Because of the screw symmetry of the body centred tetragonal
structure of the 122 materials, the in-plane elliptical shape of the Fermi surface at the corner of the zone
 rotates by $\pi/2$ between the centre and top of the zone (see Fig.\ \ref{Fig:BaFe2P2_Slices}).  This type
of distortion has been described as resembling a snake that has swallowed a chain \cite{BergemannMJFO03}.   Exactly
half way up the zone, this odd symmetry $\cos ck_z\sin 2\phi$ term ($k_{2,1}$ in the notation of a cylindrical harmonic
expansion \cite{BergemannMJFO03,RourkeBBMCCH10}) is zero, and the shape is closest to circular (even symmetry cosine
distortion terms remain) (see Fig.\ \ref{Fig:BaFe2P2_Slices}).  The anisotropy of the individual Fermi surface sheets,
parameterized by the anisotropy of the plasma frequencies (which in the isotropic scattering rate limit is equal to the
square-root of the resistance anisotropy) are shown in Table \ref{Table:plamsa}.

\begin{table}
\caption{Mass enhancement factors ($\lambda=m^*/m_b-1$) in iron-phosphides. Here $m^*$ is the measured thermal mass and
$m_b$ is the DFT band structure mass at the same angle for with the band shifted to best match the observed dHvA
frequency. For Ba-122, Sr-122 and LaFePO the orbit labels are as in Figs.\ (\ref{Fig:BaSrFe2P2_Rot}) and
(\ref{Fig:LaFePO_Rotation}). For Ca-122 the hole orbit labelled here as $\delta_2$ is the $\beta$ orbit in Fig.\
(\ref{Fig:CaFe2P2_Rot}). }\flushright
\begin{tabular}{lllll}
\hline
\hline
&LaFePO & BaFe$_2$P$_2$& SrFe$_2$P$_2$&CaFe$_2$P$_2$\\
Electron $\alpha_1$&  1.3 &0.8 &1.1 &0.5 \\
Electron $\alpha_2$&  1.6 &0.9 &0.9 &0.5 \\
Electron $\beta_1$ &  1.0 &0.7 &0.6 &- \\
Electron $\beta_2$ &  1.1 &0.8 &0.6 &- \\
Hole $\delta_1$    &  1.4 & -  & -  &- \\
Hole $\delta_2$    &  -   &0.8 &0.7 &0.5\\
Hole $\gamma_1$    & -    &0.6 &0.5 &-\\
Hole $\gamma_2$    & -    &0.6 &0.7 &-\\
\hline
\hline
\end{tabular}
\label{Table:massenhacements}
\end{table}

For SrFe$_2$P$_2$ the calculated ($\alpha/\beta$) electron Fermi surfaces are brought into almost perfect alignment
with the data by shifting them up by 49\,meV and 59\,meV respectively.  The outer hole sheet in this material has very
strong warping close to the top of the zone which as in LaFePO is caused by the band acquiring strong Fe $d_z^2$
character. At an angle of $\theta \sim 50^\circ$ the large warped sections of Fermi surface are almost parallel and so
there the curvature factor strongly increases. Hence, it is close to this angle where this orbit is most easily
observed. Interestingly, the calculation correctly predicts the size of this part of this Fermi surface without any
band energy shifting necessary.  As will be discussed below this likely is because this Fermi surface section is does
not nest with any of the electron sheets.  The inner hole sheet is somewhat smaller than the electron sheets and is
shifted down in energy by 110\,meV compared to the DFT calculation.  This causes the sheet to pinch off (see Fig.\
\ref{Fig:122_FS}), so $\delta_1$ in Fig.\ \ref{Fig:BaSrFe2P2_Rot} ceases to exist.

\begin{table}
\caption{Calculated in-plane ($\omega^*_{p,a}$) and $c$-axis ($\omega^*_{p,c}$) plasma frequencies in eV units using
the bands shifted to best fit the dHvA data.  The values have been renormalized by factor $(m^*/m_b)^{-1/2}$ where
$m^*/m_b$ is the average mass enhancement for the particular sheet (see Table \ref{Table:massenhacements}).  The
anisotropy $\Gamma=\omega^*_{p,a}/\omega^*_{p,c}$, which in the isotropic scattering approximation is the square root
of the conductivity anisotropy,  is also given for each Fermi surface sheet.  For LaFePO the possible contribution of
the small spherical $d_{z^2}$ sheet has been ignored. The anisotropy of the remaining two hole sheets is a lower bound
as all orbits have not clearly yet been identified.} \flushright
\begin{tabular}{lllllllllllll}
\hline
\hline
& \multicolumn{3}{c}{LaFePO} & \multicolumn{3}{c}{BaFe$_2$P$_2$} &  \multicolumn{3}{c}{SrFe$_2$P$_2$} &\multicolumn{3}{c}{CaFe$_2$P$_2$}\\
&$\omega_{p,a}^*$ & $\omega_{p,c}^*$ & $\Gamma$&$\omega_{p,a}^*$ & $\omega_{p,c}^*$ & $\Gamma$&$\omega_{p,a}^*$ & $\omega_{p,c}^*$ & $\Gamma$&$\omega_{p,a}^*$ & $\omega_{p,c}^*$ & $\Gamma$\\
Hole ($\delta$)&0.55   &0.06   &8.7    &0.68    &0.28    &2.4    &0.54    &0.26    &2.1    &1.50    &2.78    &0.5\\
Hole ($\gamma$)&0.55   &0.09   &6.2    &1.02    &1.06    &1.0    &1.22    &1.70    &0.7     &       &        &    \\
Elec ($\alpha$)&0.66   &0.03   &26     &0.98    &0.42    &2.3    &1.28    &0.25    &5.2    &1.86    &1.42    &1.3\\
Elec ($\beta$) &1.04   &0.20   &5.3    &1.32    &0.25    &5.3    &1.27    &0.40    &3.2    &        &        &    \\
\hline
\hline
\end{tabular}
\label{Table:plamsa}
\end{table}

An important check of the band shifting fitting of the DFT calculations to the dHvA data is to calculate the volume of
the Fermi surface sheets.  These materials are all expected to be compensated metals, so they should have equal numbers
of electrons and holes.  For SrFe$_2$P$_2$ with the shifts, the $\delta$, $\gamma$, $\alpha$ and $\beta$ bands contain
-0.026, -0.284, 0.108 and 0.197 hole/electrons respectively, and so the charge balance remains almost exact. The
imbalance of 0.005 extra holes per unit cell is equivalent to $<2$\% of the total volume of the hole sheets.

The data for BaFe$_2$P$_2$ are in many ways quite similar to those for SrFe$_2$P$_2$.  The general shape of all four
sheets of Fermi surface are very similar and also the absolute sizes are quite comparable.   The most significant
difference is the size of the inner hole sheet, particularly orbit $\delta_2$.  The cross section of this sheet is
almost identical to the $\alpha$ electron sheet. In fact,  the frequency $\delta_2$ is right between the two
extremities of that from the $\alpha$ sheet.  This suggest the possibility that there could be almost perfect nesting
between these electron and hole sheets. As mentioned above, however the electron sheets have quite elliptical
cross-sections at the centre and top of the zone so although the areas match the shape does not. However, half way
between these two extremities the electron sheet becomes almost circular and therefore matches very well the hole sheet
(see Fig.\ \ref{Fig:BaFe2P2_Slices}).  In addition the band-character of these sheets is also very similar.  It should
be recalled that despite such good nesting BaFe$_2$P$_2$ is neither superconducting nor magnetic.  Hence this is a good
example showing that nesting is a necessary but not a sufficient condition for the formation of those ground states.
Many body correlation effects which become stronger for the arsenide materials clearly also play a key role.

The band shifts needed to get the DFT calculation to agree with the data for BaFe$_2$P$_2$ (+68/+58\,meV for the
inner/outer electron sheets and -113\,meV for the inner hole sheet) are almost identical to those for SrFe$_2$P$_2$. In
particular, again it is found that the highly warped $d_{z^2}$ section of the outer hole sheet agrees well with the
calculation without any shift needed.

\begin{figure*}
\flushright
\includegraphics*[width=13cm]{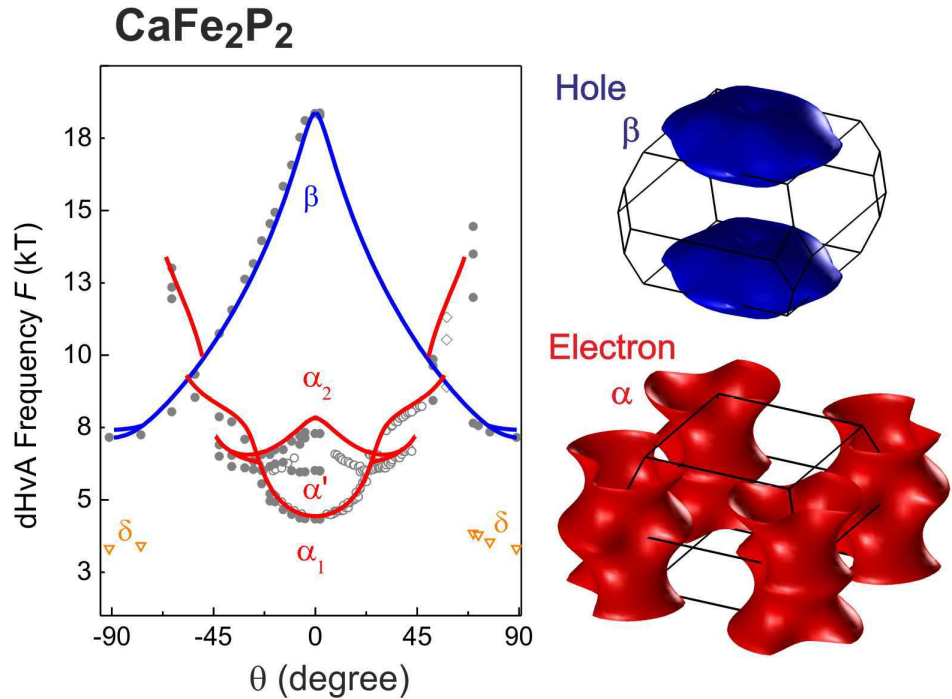}
\caption{Experimental dHvA frequencies versus magnetic field angle for rotations from $B\|c$ (001) to $B\|a$ (100) for
CaFe$_2$P$_2$ \cite{Coldea09}.  The electron and hole Fermi surface are shown to the right of the figure. Figure
adapted from Ref.\ \cite{Coldea09}.} \label{Fig:CaFe2P2_Rot}
\end{figure*}

Next we turn to CaFe$_2$P$_2$.  Here the dHvA spectra are significantly simpler than the other 122 compounds with far
fewer observed frequencies (see Fig.\ \ref{Fig:CaFe2P2_Rot}).  As remarked above (Fig.\ \ref{Fig:122_FS}), the Fermi
surface of this material is markedly different from the other 122s.  The hole Fermi surface has changed from two quasi
two dimensional cylinders to a single flat pancake which is depressed close to the $\Gamma-Z$ line.  The electron
sheets have also become much more warped but still roughly keep their cylindrical shape.  For this material, it is
found that the DFT calculations agree almost perfectly with the dHvA data  without any adjustment of the band energies
(to with $\pm$ 10\,meV). This is an important counter-example to the results found for LaFePO, BaFe$_2$P$_2$ and
SrFe$_2$P$_2$.  It would therefore appear to be clear that the band shifts are not an inherent inaccuracy in the
mean-field band-structure but are rather caused by non-mean-field correlation effects which act between almost nested
Fermi surfaces. A obvious candidate for this are spin-fluctuations.

The mass enhancements found from the dHvA measurements for all three 122s plus LaFePO are shown in Table
\ref{Table:massenhacements}. CaFe$_2$P$_2$ has the smallest enhancements, which are about two times larger than those
expected from conventional electron-phonon coupling ($\lambda_{EP}\simeq 0.23$)\cite{Yildirim09}.  For SrFe$_2$P$_2$
the enhancements are quite sheet dependent, being only slightly larger than for the Ca compound for most sheets except
for the most two dimensional $\alpha$ electron sheets where they are a factor of two larger.  BaFe$_2$P$_2$ again has
some sheet dependence and the average $\lambda=0.74$ -- about 50\% larger than for CaFe$_2$P$_2$.  A useful check of
consistency is to apply the measured mass enhancements for each sheet to the calculated density of states for each
sheet and therefore get an estimate of the electronic specific heat.  For SrFe$_2$P$_2$ this gives
$\gamma_E=10.4(2)$\,mJ mol$^{-1}$K$^{-2}$, which agrees well with the experimental value of $\gamma_E=11.6(2)$\,mJ
mol$^{-1}$K$^{-2}$ \cite{Caroline11}.

\subsection{BaFe$_2$(As$_{1-x}$P$_x$)$_2$}
Although studies of the non-superconducting end members described above can give valuable insight into the bulk
electronic structure of the superconductors, it is clearly preferable to measure the actual superconducting materials
themselves.  Unfortunately, in the main, the condition that the samples must be free of disorder means that it is not
possible to observe quantum oscillations in heavily doped superconducting samples.  This is compounded by the fact that
often the highest $T_c$ samples also have very high $H_{c2}$ values and so the normal state cannot be accessed.  LaFePO
and KFe$_2$As$_2$ \cite{Terashima10} are rare examples where superconductivity occurs in the (close to) stoichiometric
composition and dHvA oscillations are observable.  However, it is unclear whether these low $T_c$ materials are
representative of the higher $T_c$ pnictides.  Fortunately, there is one exception to this. In the isovalently
substituted BaFe$_2$(As$_{1-x}$P$_x$)$_2$ series the residual resistance, and hence scattering, remains low even as the
maximum $T_c$ is approached ($T_c^{max}$=30\,K for $x=0.33$) \cite{Kasahara10}.  In this material, and other pnictides,
there is clear evidence that the transport properties develop `non-Fermi liquid' properties as the maximum $T_c$ is
approached.  The resistivity follows a power law as a function of temperature, $\rho \propto T^n$ where the exponent
$n$ evolves from its conventional Fermi liquid value of 2 for the overdoped phosphide to close to 1 at optimal doping.
Such behaviour is strongly reminiscent of the high $T_c$ cuprate superconductors \cite{Kubo91} and heavy Fermions in
the vicinity of a quantum critical point \cite{Custers03}.  In BaFe$_2$(As$_{1-x}$P$_x$)$_2$ the highest $T_c$ occurs
approximately where the SDW order temperature extrapolates to zero and hence is likely strongly influenced by the
presence of a quantum critical point \cite{Dai09}.

\begin{figure*}
\flushright
\includegraphics*[width=13cm]{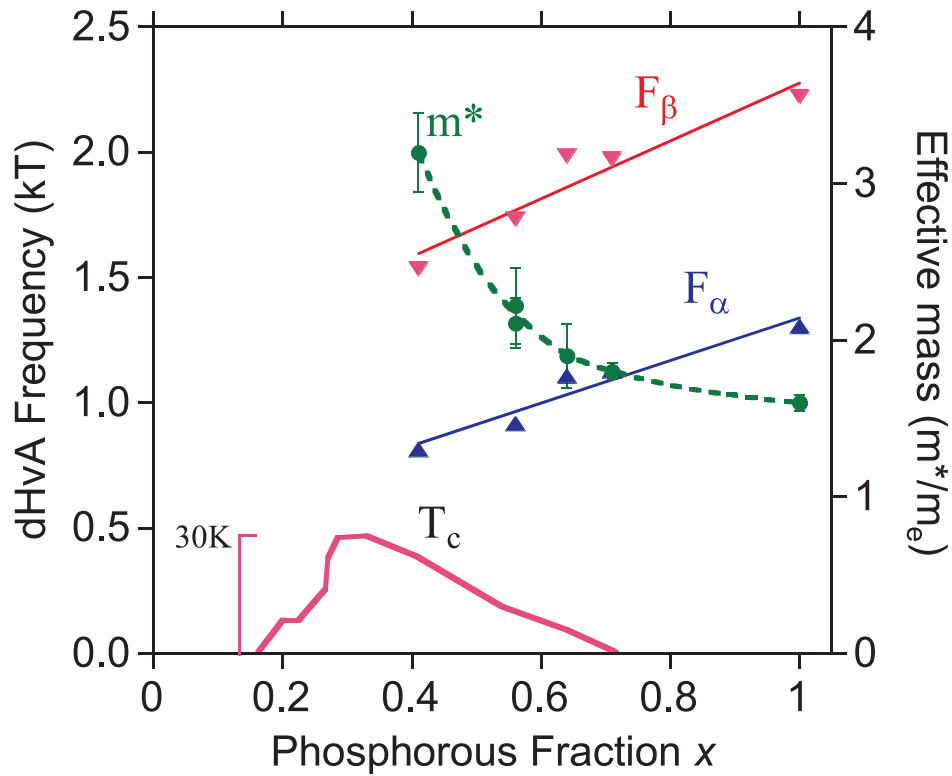}
\caption{Experimental dHvA frequencies for the $\alpha$ and $\beta$ electron orbits of BaFe$_2$(As$_{1-x}$P$_x$)$_2$
versus the phosphorous content $x$.  Where more than one $\alpha/\beta$ frequency was observed the average
frequency/mass is plotted here. The $x$ dependence of the effective mass $m^*$ of the $\beta$ orbits and $T_c$ is also
shown. Figure adapted from Ref.\ \cite{Shishido10,Arnold11}. The more accurate value of $m^*$ from Ref.\
\cite{Arnold11} for $x=1$ is plotted in preference to that in Ref.\ \cite{Shishido10}.} \label{Fig:BaFe2PAS2_Allvsx}
\end{figure*}

Using pulsed magnetic fields up to 60\,T dHvA oscillations were observed in the BaFe$_2$(As$_{1-x}$P$_x$)$_2$ series
for $x$ down to 0.42, where $T_c$ has reached 25\,K which is $\sim$ 80\% of the maximum \cite{Shishido10}.  The quite
remarkable fact that oscillations could still be seen in such a heavily substituted composition implies that As/P cross
substitution induces very little disorder on the conducting Fe plane.  The observed frequencies were identified as
coming from the electron sheets, and for the lower values of $x$ (higher $T_c$) only a single frequency from each
electron band was discernable.  In the higher $T_c$ samples oscillations were only observable at the highest fields and
hence the Fourier peaks are too broad to see the individual peaks.

Two important results were found.  First it was observed that both the $\alpha$ and $\beta$ frequencies decrease
markedly with decreasing $x$ (see Fig.\ \ref{Fig:BaFe2PAS2_Allvsx}).  No such decrease is expected from the DFT band
structure calculations. Second, the effective mass increases strongly as the maximum $T_c$ is approached.  It is likely
that both these changes are caused by the increase in correlations (spin fluctuations) as the SDW ordered state is
approached. It is not clear from the available data whether the increase in effective mass is a divergence associated
with a possible quantum critical point or rather just a general increase with decreasing $x$. Unfortunately, no data is
yet available for the values of $x$ below the critical value.

Recently  Analytis et al. \cite{Analytis10} reported measurements for the $x = 0.63$ with $T_c$=7\,K where orbits on
the smaller hole sheet were observed. The frequency of these orbits were found to correspond to almost the same
cross-sectional area as the smaller electron pocket, very similar to that found for the $x=1$ composition described
above.  Hence it would appear that the \textit{relative} size of the electron and hole sheets remain remarkably
constant as a function of $x$. As the electron sheets shrink with decreasing $x$ the hole bands shrink by a similar
amount without any major change in topology.  This is supported by recent ARPES results for the $x = 0.38$ composition
\cite{Yoshida11} which show that in this close to optimal doped composition the shape of the Fermi surfaces closely
resemble the end member BaFe$_2$P$_2$.

A semi quantitative model of the electron / hole pocket shrinking of the quasi-nested electron-hole Fermi surfaces of
iron-pnictides has been suggested by Ortenzi et al.\ \cite{OrtenziCBP09}.  Here scattering caused by coupling of the
itinerant electrons to a bosonic spin fluctuation mode (assumed to be Einstein-like) causes both electron and hole
pockets to shrink and also the effective mass (and $T_c$) to increase.  This was applied to a simplified model of the
LaFePO Fermi surface and reasonable quantitative agreement was found between the magnitude of the energy shifts (which
in this model determines the strength of the scattering potential) and the mass enhancement and even $T_c$. It should
be noted however that taken at face value this model would predict low-energy correlation-driven renormalisation of the
band energies (i.e., kinks in $\varepsilon(\bm{k})$ close to the Fermi level), whereas a comparison of the dHvA data to
ARPES suggests the renormalisation occurs over a much wider energy range.  It would however be intriguing to see if an
extension of this model could explain the correlations between band energy shifts, effective masses and $T_c$ in the
BaFe$_2$(As$_{1-x}$P$_x$)$_2$ series.  This could prove to be a decisive test of the spin fluctuation pairing model in
the iron-pnictides.

\section{Summary and Conclusions}

Quantum oscillation measurements have proven to be a very valuable probe of the bulk electronic structure of iron-based
superconductors.   Combined with first principle electron structure calculations a very detailed model of the Fermi
surface and the quasi-particle dynamics can be obtained. This review has shown that density functional theory
calculations provide a good starting point.  Using $k$-independent shifts to the band energies the band structure
calculations can be fitted to the data using very few parameters.  These shifts appear to originate from strong
electron correlations and are present even for the non-superconducting, non-magnetic phosphides. Importantly, it is
found that only those parts of the Fermi surface where there is electron-hole quasi-nesting are shifted in energy. For
CaFe$_2$P$_2$ no shifting is necessary.   In the BaFe$_2$(As$_{1-x}$P$_x$)$_2$ series the shifting increases as the SDW
phase transition is approached and is accompanied by a strong increase in the effective mass.   It is hoped that in the
near future, more sophisticated first principle calculations which go beyond the mean-field level, such as dynamical
mean field theory, may explain these results and lead to new insights into the physic of iron-based superconductors.

\section{Acknowledgements}
I am indebted to my many colleagues who have contributed to many of the experiments described in this review.   In
particular, I would like to acknowledge the contributions of Amalia Coldea, Jon Fletcher, Brendan Arnold, Caroline
Andrew, Ali Bangura, Patrick Rourke (Bristol), James Analytis, Ian Fisher (Stanford), Yuji Matsuda, Takasada Shibauchi,
Shigeru Kasahara (Kyoto), Cyril Proust, David Vignolles, and Baptiste Vignolle (Toulouse) with whom I personally have
worked most closely. This work was supported by the UK EPSRC.

\section{References}
\bibliographystyle{ioppart-num}
\bibliography{ROPIP_QOReview_2011}

\providecommand{\newblock}{}
\begin{thebibliography}{10}
\expandafter\ifx\csname url\endcsname\relax
  \def\url#1{{\tt #1}}\fi
\expandafter\ifx\csname urlprefix\endcsname\relax\def\urlprefix{URL }\fi
\providecommand{\eprint}[2][]{\url{#2}}

\bibitem{SinghD08}
Singh D~J and Du M~H 2008 {\em Phys. Rev. Lett.\/} {\bf 100} 237003

\bibitem{MazinSJD08}
Mazin I~I, Singh D~J, Johannes M~D and Du M~H 2008 {\em Phys. Rev. Lett.\/}
  {\bf 101} 057003

\bibitem{KurokiOAUTKA08}
Kuroki K, Onari S, Arita R, Usui H, Tanaka Y, Kontani H and Aoki H 2008 {\em
  Phys. Rev. Lett.\/} {\bf 101} 087004

\bibitem{ChubukovEE08}
Chubukov A~V, Efremov D~V and Eremin I 2008 {\em Phys. Rev. B\/} {\bf 78}
  134512

\bibitem{CvetkovicT09}
Cvetkovic V and Tesanovic Z 2009 {\em EPL\/} {\bf 85} 37002

\bibitem{KamiharaHHKYKH06}
Kamihara Y, Hiramatsu H, Hirano M, Kawamura R, Yanagi H, Kamiya T and Hosono H
  2006 {\em J. Am. Chem. Soc.\/} {\bf 128} 10012

\bibitem{KamiharaWHH08}
Kamihara Y, Watanabe T, Hirano M and Hosono H 2008 {\em J. Am. Chem. Soc.\/}
  {\bf 130} 3296--3297

\bibitem{MazinS09}
Mazin I~I and Schmalian J 2009 {\em Physica C\/} {\bf 469} 614--627

\bibitem{Shoenberg}
Shoenberg D 1984 {\em Magnetic Oscillations in Metals\/} (Cambridge: Cambridge
  University Press)

\bibitem{Shishido10}
Shishido H, Bangura A~F, Coldea A~I, Tonegawa S, Hashimoto K, Kasahara S,
  Rourke P~M~C, Ikeda H, Terashima T, Settai R, Onuki Y, Vignolles D, Proust C,
  Vignolle B, Mccollam A, Matsuda Y, Shibauchi T and Carrington A 2010 {\em
  Phys. Rev. Lett.\/} {\bf 104} 057008

\bibitem{Rotter08}
Rotter M, Tegel M, Johrendt D, Schellenberg I, Hermes W and P\"ottgen R 2008
  {\em Phys. Rev. B\/} {\bf 78} 020503

\bibitem{Tanatar10}
Tanatar M~A, Blomberg E~C, Kreyssig A, Kim M~G, Ni N, Thaler A, Bud'ko S~L,
  Canfield P~C, Goldman A~I, Mazin I~I and Prozorov R 2010 {\em Phys. Rev. B\/}
  {\bf 81} 184508

\bibitem{ChuAPDLYF10}
Chu J~H, Analytis J~G, Press D, De~greve K, Ladd T~D, Yamamoto Y and Fisher I~R
  2010 {\em Phys. Rev. B\/} {\bf 81} 214502

\bibitem{ChuADMIYF10}
Chu J~H, Analytis J~G, De~greve K, Mcmahon P~L, Islam Z, Yamamoto Y and Fisher
  I~R 2010 {\em Science\/} {\bf 329} 824--826

\bibitem{Yi2009}
Yi M, Lu D~H, Analytis J~G, Chu J~H, Mo S~K, He R~H, Hashimoto M, Moore R~G,
  Mazin I~I, Singh D~J, Hussain Z, Fisher I~R and Shen Z~X 2009 {\em Phys. Rev.
  B\/} {\bf 80} 174510

\bibitem{AnalytisMCRBKJF09}
Analytis J~G, Mcdonald R~D, Chu J~H, Riggs S~C, Bangura A~F, Kucharczyk C,
  Johannes M and Fisher I~R 2009 {\em Phys. Rev. B\/} {\bf 80} 064507

\bibitem{SebastianGHLSML08}
Sebastian S~E, Gillett J, Harrison N, Lau P~H~C, Singh D~J, Mielke C~H and
  Lonzarich G~G 2008 {\em J. Phys. Cond. Matt.\/} {\bf 20} 422203

\bibitem{HarrisonMMBRT09}
Harrison N, Mcdonald R~D, Mielke C~H, Bauer E~D, Ronning F and Thompson J~D
  2009 {\em J. Phys.-Condes. Matter\/} {\bf 21} 322202

\bibitem{harmonic}
Note that Terashima \textit{et al.}, suggest that the second of these
  frequencies is actually the second harmonic of the first
  \cite{Terashima1103.3329}

\bibitem{Terashima1103.3329}
Terashima T, Kurita N, Tomita M, Kihou K, Lee C~H, Tomioka Y, Ito T, Iyo A,
  Eisaki H, Liang T, Nakajima M, Ishida S, Uchida S, Harima H and Uji S
  ArXiv:1103.3329

\bibitem{ColdeaFCABCEFHM08}
Coldea A~I, Fletcher J~D, Carrington A, Analytis J~G, Bangura A~F, Chu J~H,
  Erickson A~S, Fisher I~R, Hussey N~E and Mcdonald R~D 2008 {\em Phys. Rev.
  Lett.\/} {\bf 101} 216402

\bibitem{Hamlin08}
Hamlin J~J, Baumbach R~E, Zocco D~A, Sayles T~A and Maple M~B 2008 {\em J.
  Phys. Cond. Mat.\/} {\bf 20} 365220

\bibitem{McqueenRWHLHWGC08}
Mcqueen T~M, Regulacio M, Williams A~J, Huang Q, Lynn J~W, Hor Y~S, West D~V,
  Green M~A and Cava R~J 2008 {\em Phys. Rev. B\/} {\bf 78} 024521

\bibitem{CarringtonCFHABACEFM09}
Carrington A, Coldea A~I, Fletcher J~D, Hussey N~E, Andrew C~M~J, Bangura A~F,
  Analytis J~G, Chu J~H, Erickson A~S, Fisher I~R and Mcdonald R~D 2009 {\em
  Physica C\/} {\bf 469} 459--468

\bibitem{SugawaraSDMKYO08}
Sugawara H, Settai R, Doi Y, Muranaka H, Katayama K, Yamagami H and Onuki Y
  2008 {\em J. Phys. Soc. Jpn.\/} {\bf 77} 113711

\bibitem{Yamaji89}
Yamaji K 1989 {\em J. Phys. Soc. Jpn.\/} {\bf 58} 1520

\bibitem{BergemannMJFO03}
Bergemann C, Mackenzie A~P, Julian S~R, Forsythe D and Ohmichi E 2003 {\em Adv.
  Phys.\/} {\bf 52} 639--725

\bibitem{BanguraRBMCHC10}
Bangura A~F, Rourke P~M~C, Benseman T~M, Matusiak M, Cooper J~R, Hussey N~E and
  Carrington A 2010 {\em Phys. Rev. B\/} {\bf 82} 140501

\bibitem{RourkeBBMCCH10}
Rourke P~M~C, Bangura A~F, Benseman T~M, Matusiak M, Cooper J~R, Carrington A
  and Hussey N~E 2010 {\em New J. Phys.\/} {\bf 12} 105009

\bibitem{wien2k}
Blaha P, Schwarz K, Madsen G~K~H, Kvasnicka D and Luitz J 2001 {\em WIEN2K, An
  Augmented Plane Wave + Local Orbitals Program for Calculating Crystal
  Properties\/} (Karlheinz Schwarz, Techn. Universit\"at Wien, Austria) iSBN
  3-9501031-1-2

\bibitem{Boeri08}
Boeri L, Dolgov O~V and Golubov A~A 2008 {\em Phys. Rev. Lett.\/} {\bf 101}
  026403

\bibitem{Jiang09}
Jiang S, Xing H, Xuan G, Wang C, Ren Z, Feng C, Dai J, Xu Z and Cao G 2009 {\em
  J. Phys. Cond. Mat.\/} {\bf 21} 382203

\bibitem{Shi10}
Shi H~L, Yang H~X, Tian H~F, Lu J~B, Wang Z~W, Qin Y~B, Song Y~J and Li J~Q
  2010 {\em J. Phys. Cond. Mat.\/} {\bf 22} 125702

\bibitem{sefat08}
Sefat A~S, Jin R, McGuire M~A, Sales B~C, Singh D~J and Mandrus D 2008 {\em
  Phys. Rev. Lett.\/} {\bf 101} 117004

\bibitem{AlirezaKGPCLS09}
Alireza P~L, Ko Y~T~C, Gillett J, Petrone C~M, Cole J~M, Lonzarich G~G and
  Sebastian S~E 2009 {\em J. Phys. Cond. Mat.\/} {\bf 21} 012208

\bibitem{AnalytisACMCMFC09}
Analytis J~G, Andrew C~M~J, Coldea A~I, Mccollam A, Chu J~H, Mcdonald R~D,
  Fisher I~R and Carrington A 2009 {\em Phys. Rev. Lett.\/} {\bf 103} 076401

\bibitem{Coldea09}
Coldea A~I, Andrew C~M~J, Analytis J~G, McDonald R~D, Bangura A~F, Chu J~H,
  Fisher I~R and Carrington A 2009 {\em Phys. Rev. Lett.\/} {\bf 103} 026404

\bibitem{Arnold11}
Arnold B, Kasahara S, Coldea A, Terashima T, Matsuda Y, Shibauchi T and
  Carrington A 2011 {\em Phys. Rev. B\/} {\bf x} x

\bibitem{Yildirim09}
Yildirim T 2009 {\em Physica C\/} {\bf 469} 425--441

\bibitem{Caroline11}
Andrew C.M.J., Ph.D. thesis, University of Bristol, (2011)

\bibitem{Terashima10}
Terashima T, Kimata M, Kurita N, Satsukawa H, Harada A, Hazama K, Imai M, Sato
  A, Kihou K, Lee C~H, Kito H, Eisaki H, Iyo A, Saito T, Fukazawa H, Kohori Y,
  Harima H and Uji S 2010 {\em J. Phys. Soc. Jpn.\/} {\bf 79} 053702

\bibitem{Kasahara10}
Kasahara S, Shibauchi T, Hashimoto K, Ikada K, Tonegawa S, Okazaki R, Shishido
  H, Ikeda H, Takeya H, Hirata K, Terashima T and Matsuda Y 2010 {\em Phys.
  Rev. B\/} {\bf 81} 184519

\bibitem{Kubo91}
Kubo Y, Shimakawa Y, Manako T and Igarashi H 1991 {\em Phys. Rev. B\/} {\bf 43}
  7875--7882

\bibitem{Custers03}
Custers J, Gegenwart P, Wilhelm H, Neumaier K, Tokiwa Y, Trovarelli O, Geibel
  C, Steglich F, Pepin C and Coleman P 2003 {\em Nature\/} {\bf 424} 524--527

\bibitem{Dai09}
Dai J~H, Si Q~M, Zhu J~X and Abrahams E 2009 {\em Proc. Natl. Acad. Sci.\/}
  {\bf 106} 4118--4121

\bibitem{Analytis10}
Analytis J~G, Chu J~H, McDonald R~D, Riggs S~C and Fisher I~R 2010 {\em Phys.
  Rev. Lett.\/} {\bf 105} 207004

\bibitem{Yoshida11}
Yoshida T, Nishi I, Ideta S, Fujimori A, Kubota M, Ono K, Kasahara S, Shibauchi
  T, Terashima T, Matsuda Y, Ikeda H and Arita R 2011 {\em Phys. Rev. Lett.\/}
  {\bf 106} 117001

\bibitem{OrtenziCBP09}
Ortenzi L, Cappelluti E, Benfatto L and Pietronero L 2009 {\em Phys. Rev.
  Lett.\/} {\bf 103} 046404

\end{thebibliography}
\end{document}